\documentclass[10pt,letterpaper,twocolumn]{article} 

\usepackage{ol2}
\usepackage[draft]{hyperref}
\usepackage{amsmath}

\begin{document}

\twocolumn[ 

\title{Illustration of Babinet's principle with ultracold atoms}


\author{Aaron Reinhard,$^{1,*}$ Jean-F\'{e}lix Riou,$^2$ Laura A. Zundel$^2$ and David S. Weiss$^{2}$}

\address{
$^1$Department of Physics, Otterbein University, 1 South Grove St., Westerville, OH 43081, USA
\\
$^2$Physics Department,  The Pennsylvania State University, 104 Davey Lab, University Park, PA 16802, USA \\
$^*$Corresponding author: areinhard@otterbein.edu
}

\begin{abstract}
We demonstrate Babinet's principle by the absorption of high intensity light from dense clouds of ultracold atoms. Images of the diffracted light are directly related to the spatial distribution of atoms. The advantages of employing Babinet's principle as an imaging technique are that it is easy to implement and the detected signal is large. We discuss the regimes of applicability of this technique as well as its limitations.
\end{abstract}


 ] 

\noindent

Babinet's principle derives from linear superposition applied to diffraction.  It says that the light field diffracted from an aperture plus the field that would be diffracted from the aperture's complement must sum to the incident light field~\cite{Babinet}.  A well-known and experimentally verified restatement of Babinet's principle says that the diffraction pattern from an opaque object is the same as that from a hole of the same shape, except for the forward beam intensity~\cite{Hecht}.  In this Letter, we demonstrate Babinet's principle using a sample of ultracold atoms as semi-opaque objects.  Light energy proportional to the integrated atom density is spontaneously scattered from the probe, which causes a dip in the electric field amplitude of the probe beam. When the undiffracted part of the probe beam is blocked, the image formed by the remaining light is the same as it would be if the electric field at the position of the atoms had the shape of the dip. Our technique amounts to imaging the diffraction pattern produced by an ``obstacle.''  The converse, or imaging the diffraction pattern of a bright ``aperture,'' is high intensity fluorescent imaging~\cite{DePue}. Along with the physical interest of applying this centuries old idea in this modern context, it has some unique features as an imaging technique.

 Imaging is the dominant way to gain information about a sample of ultracold atoms.  Images of atom clouds may be taken either in-situ or after ballistic expansion.  They can be used to access a wide variety of information about an evolving cloud of ultracold atoms, including position and momentum distributions~\cite{Nelson, Bakr, Sherson, Kinoshita}, coherence properties~\cite{Greiner, Albiez}, and quasimomentum distributions~\cite{Greiner2}. Techniques for imaging ultracold atom clouds can be placed into three broad categories: fluorescent, absorptive, and dispersive.  For imaging low density clouds, on-resonance fluorescence or absorption with low probe intensities are often employed.  These techniques are not easy to use for dense atomic clouds,  since probe beam absorption makes atoms see a nonuniform probe intensity along the direction of the beam.  Low intensity absorption imaging of small clouds further suffers from distortion of the image due to ``lensing'' caused by the atoms' index of refraction and the cloud's small radius of curvature.

Atomic clouds with optical thickness much larger than one, such as in-situ Bose-Einstein condensates (BECs) or degenerate Fermi gases, are often imaged using dispersive techniques~\cite{Andrews, Andrews2, Turner}.  These techniques rely on the interference of phase shifted probe light with unshifted light.  Because the probe is highly detuned from resonance, these techniques do not suffer significantly from image distortion due to lensing or from probe saturation.  Dispersive imaging is often the most attractive technique for imaging dense clouds, although it does require a highly detuned (hundreds of atomic linewidths) probe beam and involves non-trivial image processing. Resonant fluorescence from a high intensity probe beam can also be used to image dense clouds~\cite{DePue}.  While fluorescent imaging is robust and requires no signal processing, it suffers from smaller signal than other imaging techniques.  This is because in fluorescent imaging, one collects a fraction of all scattered light, while in other techniques one collects the entire available signal which is imprinted on the probe beam.

The Babinet's principle technique we describe here uses absorption of a high-intensity resonant probe to image dense clouds.  We have used it to study the self-trapping dynamics of quasi-1D gases in a 2D optical lattice~\cite{Reinhard}.  Its features are that it is simple to implement, has high signal to noise, and requires minimal image processing compared to dispersive imaging. However, it requires a very high intensity probe and involves a reduction of the resolution by a factor $\sqrt{2}$ and a small amount of density-dependent distortion at very high densities (although the distortion is smaller than for low-intensity absorption imaging).

\begin{figure}[htp]
\centerline{ \scalebox{0.27} {\includegraphics{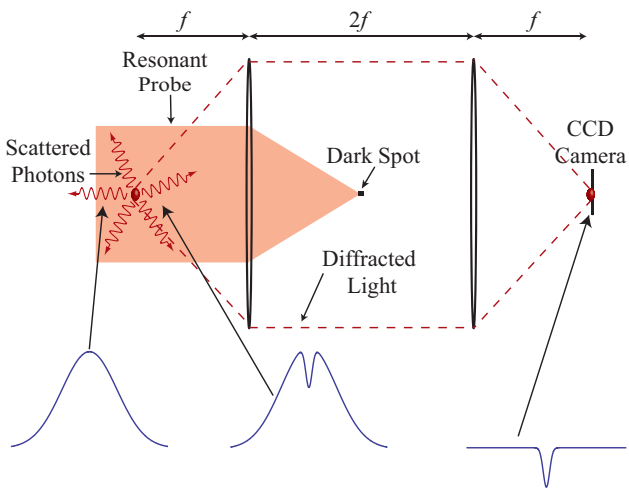}}} \caption{ \label{Fig_ImgSys} Imaging system used for high intensity absorption imaging, where $f$ is the focal length of the lenses.  Photons are resonantly scattered out of the probe beam by the atom cloud.  Diffracted probe light within the cone illustrated by dotted lines is imaged on the camera, while the incident beam is blocked by a dark spot.  At the bottom of the figure, we show the transverse electric field profile of the probe beam at various points in the imaging system.  The size of the cloud relative to the beam waist has been exaggerated for clarity.}
\end{figure}

In our experiment, a resonant probe beam of intensity $I_o$ and $1/e^2$ waist 1.3~mm impinges on a $^{87}$Rb BEC and continues through the imaging system shown schematically in Fig.~\ref{Fig_ImgSys}. When $I_{\rm{o}} - OT \times I_{\rm{s}} \gg 1$, where $I_{\rm{s}}$ is the saturation intensity and $OT$ is the optical thickness, all the atoms spontaneously emit at the maximum rate, and  the probe intensity is decreased by an amount proportional to the integrated atom density along a line.  Because the light intensity is so far above saturation, the ground and excited state populations are nearly equal, which should suppress non-linear effects~\cite{Labeyrie}.  We  block the undiffracted part of the probe using a $400~\mu$m diameter ``dark spot'' in the focal plane, and collect the diffracted light with a one-to-one imaging system. We note that our Babinet's principle technique is different from previously reported high intensity absorption imaging in which the ``standard'' absorption imaging geometry was used with probe beams about 40 times less intense than the ones used here~\cite{Reinaudi}.  Since the probe beam had to be filtered before the camera to prevent saturation, the per atom signal was reduced as the OT went up. It is unchanged in the Babinet's principle technique.

We now derive the light intensity at the output of our imaging system as a function of the 2D integrated atomic density.  We denote the electric field of the incident probe beam as $\vec{E}_{\rm{o}}$, the decrease in electric field due to spontaneous emission as $\vec{E}_{\rm{dip}}$, and the field transmitted through the atoms as $\vec{E}_{\rm{prop}}$.  By the superposition principle,
\begin{equation} \label{Efield_superposition}
    \vec{E}_{\rm{prop}} = \vec{E}_{\rm{o}} - \vec{E}_{\rm{dip}} \quad .
\end{equation}
By conservation of energy,
\begin{equation} \label{Icons}
    I_{\rm{prop}} = I_{\rm{o}} - I_{\rm{scat}} \quad .
\end{equation}
where $I_{\rm{scat}}$ is the part of the incident intensity that is spontaneously scattered out of the probe beam, and $I_{\rm{prop}}$ is the intensity of the probe beam immediately after passing through the atoms. Note that only in the limit of a completely black absorber, where $I_{\rm{prop}}=0$, is $I_{\rm{scat}}$ proportional to $|\vec{E}_{\rm{dip}}|^2$. It is easy to show that
\begin{equation} \label{Is}
    I_{\rm{scat}} = I_{\rm{s}} \sigma_{\rm{o}} n_{\rm{2D}}(x,y) \quad ,
\end{equation}
where $I_{\rm{s}}$ is the saturation intensity, $\sigma_{\rm{o}}$ is the resonant scattering cross section, and $n_{\rm{2D}}(x,y)$ is the 3D density distribution integrated along the line of sight.

Using $I=\frac{c\epsilon_{\rm{o}}}{2} \vert \vec{E} \vert^{2}$ in free space, we can combine Eqs.~\ref{Efield_superposition}~and~\ref{Icons} to obtain
\begin{equation} \label{Eatoms}
    E_{\rm{dip}} = E_{\rm{o}} - \sqrt{ E_{\rm{o}}^{2} - \frac{2}{c\epsilon_{\rm{o}}} I_{\rm{scat}}} \quad .
\end{equation}
Since $I_{\rm{o}} \gg I_{\rm{s}}$, $E_{\rm{o}}^{2} = \frac{2}{c\epsilon_{\rm{o}}} I_{\rm{o}} \gg \frac{2}{c\epsilon_{\rm{o}}} I_{\rm{scat}}$ and we may expand the square root to first order to obtain
\begin{equation} \label{Eatoms_hiabs}
    E_{\rm{dip}} \approx  \frac{1}{c\epsilon_{\rm{o}}} \frac{I_{\rm{scat}}}{E_{\rm{o}}} \quad .
\end{equation}
Because of the dark spot, only $ - \vec{E}_{\rm{dip}}$ propagates through the entire imaging system. We detect a bright image on the CCD due to the dip in the probe electric field caused by light being scattered out of the probe beam. This is the essence of Babinet's principle, that a dip in the electric field of a beam gives the same diffraction pattern as an electric field with the shape of the dip. Specifically, the intensity detected on the CCD is
\begin{equation} \label{Iatoms_hiabs}
    I_{\rm{sig}} \approx \frac{1}{4} \frac{I_{\rm{scat}}^{2}}{I_{\rm{o}}} \quad .
\end{equation}

For the probe to not get depleted, $I_o$ must be at least several times  bigger than $I_{\rm{scat}}$ for every path through the atoms. The largest obtainable $I_{\rm{sig}} \approx I_{\rm{scat}}/4$, which for our 0.1 numerical aperture imaging lens is 177 times larger than the peak signal from collecting fluorescence.  For a given choice of $I_o$, $I_{\rm{sig}}$ is proportional to $n_{\rm{2D}}^2(x,y)$. Since $n_{\rm{2D}}^2(x,y)$ features are narrower than $n_{\rm{2D}}(x,y)$ features, the effective resolution of the imaging system is reduced by about a factor of $\sqrt{2}$. From Eq.~\ref{Iatoms_hiabs}, it is clear that a quantitative measurement of the total atom number requires a quantitative measurement of $I_{\rm{o}}$.

The linearly polarized probe beam is tuned to resonance with the $^{87}$Rb D2 transition, $ \vert F=2, m_{\rm{F}} \rangle \rightarrow \vert F'=3, m_{\rm{F}} \rangle$ with $I_{\rm{o}}$ up to $600$ times $I_{\rm{s}}$. To avoid loss to the $ \vert F=1, m_{\rm{F}} \rangle$ state via off-resonant scattering from $\vert F'=2, m_{\rm{F}} \rangle$ we repump on the $\vert F=1, m_{\rm{F}} \rangle \rightarrow \vert F'=2, m_{\rm{F}}' \rangle$ transition using a $10\%$ sideband.  The atoms therefore spend little time in the $\vert F=1, m_{\rm{F}} \rangle$ state. Also, any light scattered out of the repumping part of the beam has the same effect on the transmitted field as light scattered on the primary transition.

\begin{figure}[htp]
\centerline{ \scalebox{0.55} {\includegraphics{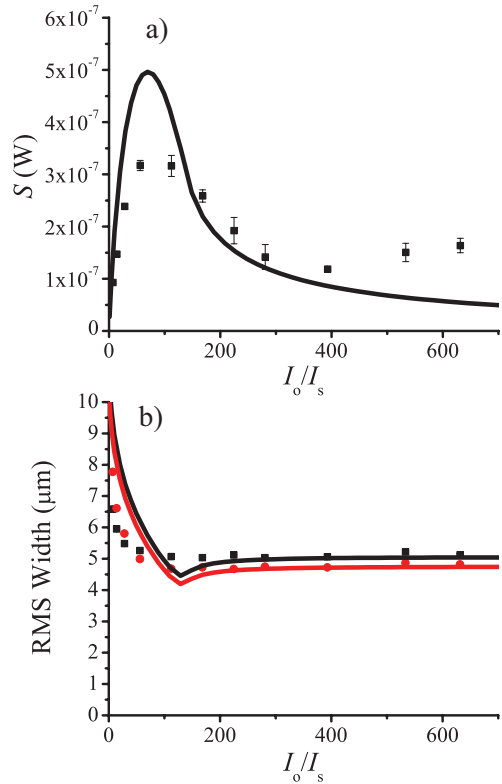}}} \caption{ \label{Fig_NumSzvsPow} (a) Integral of the detected intensity distribution, $S$, as a function of probe intensity for a Bose Einstein condensate of peak optical thickness 146 ($I_{\rm{s}}$ is the saturation intensity). Data (squares) is shown along with the results of a simulation with no free parameters (solid line).  (b) RMS width of the detected intensity distribution.  The black squares and line are for the vertical direction and the red circles and line are for the horizontal direction.}
\end{figure}

Figure~\ref{Fig_NumSzvsPow}a shows the spatial integral of the detected intensity distribution, $S$, for an in-situ BEC image.  In the limit of small probe power, the signal intensity increases with power as the probe beam intensity exceeds saturation everywhere and more atoms scatter photons.  In the opposite limit, Eq.~\ref{Iatoms_hiabs} is valid and the signal intensity decreases with probe power.  In between, there is a maximum.  Figure~\ref{Fig_NumSzvsPow}b shows the measured root-mean-square (RMS) widths.  In the large intensity limit, where each atom's absorption is totally saturated, the measured RMS width is independent of the probe beam intensity.

To quantitatively understand the behavior of these data, we have performed a numerical simulation of our system. Using the exact expression for the atomic susceptibility, $\chi$, we first determine the change in amplitude and phase of a resonant optical field after it passes through an atomic cloud. We then simulate the propagation of that field through our imaging system by numerically solving the Fresnel diffraction integrals.  The calculation incorporates an average over the measured time dependence of our $3~\mu$s probe pulse, which rises and falls in 210~ns.  Accounting for the known time dependence has up to a 55~$\%$ effect on $S$ at large probe intensity (bringing the curve closer to the data), but a negligible effect on the predicted widths.  The inputs to the calculation are the atom number, $N=2.7 \times 10^5$ determined from low-intensity absorption imaging after the cloud is allowed to expand, as well as the in-situ size of the BEC, inferred from the data in Fig.~\ref{Fig_NumSzvsPow}b and the known imaging system resolution of $1.7~\mu$m.  Because the measured atomic distributions are well fit by gaussians, the calculation uses a gaussian input cloud, with RMS widths of $6.51~\mu$m in the horizontal direction and $6.13~\mu$m in the vertical direction.  The results of our simulation are insensitive to the distribution of atoms among Zeeman sublevels.  This is because in the limit of high probe intensity, a reduction in the Clebsch-Gordan coefficient for the transition will lead to a larger saturation intensity, but a smaller optical thickness for the cloud, and these two effects approximately cancel.

We find good agreement between the theory and experiment for the horizontal RMS width.  Note that, for this comparison, we are not accounting for the small effect of density-dependent probe beam distortion (see below).  The relative minimum in the theoretical curve results from the transition from a saturated distribution to a gaussian as the probe intensity is increased and the beam remains well above saturation everywhere, a shape transition not seen in the experiment.  The theoretical curve of $S$ vs probe intensity has the same qualitative behavior as the data.  In the limit of large intensity the simulation drops as approximately $I_{\rm{o}}^{-1}$ in accord with Eq.~\ref{Iatoms_hiabs}.  The data, however, reaches a smaller peak value, decreases more slowly, and does not fall below about 1/3 of its peak value. Collection of spontaneously emitted light does not account for this discrepancy.  That the theory describes the data as well as it does with no free parameters, leaves little doubt that the essential physics here is that of Babinet's principle.

Figure~\ref{Fig_NumSzvsPow} suggests the appropriate strategy for implementation of this imaging technique.  One should take images of the atom cloud as a function of probe intensity, and then choose an intensity large enough that the size of the image is independent of intensity, but small enough to give favorable signal to noise.  The probe intensity that gives the peak signal decreases as the density of the atomic cloud becomes smaller and the sample becomes less optically thick.

Our method is valid in the limit of a small dark spot, which only filters the low spatial frequencies comprising $E_{\rm{o}}$, not the higher spatial frequencies comprising $E_{\rm{dip}}$.  The dark spot should be large enough to completely block the propagation of the field $\vec{E}_{\rm{o}}$ in the Fourier plane, but small compared to the spatial extent of the field $\vec{E}_{\rm{dip}}$.  For our $400~\mu$m diameter dark spot, our calculations show distortions of the image for a cloud of RMS width greater than $\sim30$~$\mu$m.  In practice, however, we can faithfully image clouds of RMS width of up to $180~\mu$m~\cite{note2}.  This is likely caused by imperfect imaging system alignment.  The location of the dark spot is chosen so that it coincides with the position of the probe beam in the Fourier plane.  If either the object or the probe beam are not exactly on the optical axis, $\vec{E}_{\rm{dip}}$ focuses to a different position in the Fourier plane than the dark spot.  Thus, for larger objects, which diffract less and make $\vec{E}_{\rm{dip}}$ smaller in the Fourier plane, the dark spot blocks only a fraction of any given low angle component of $\vec{E}_{\rm{dip}}$.

Because atoms in the excited state have the opposite dispersion of atoms in the ground state, and the atoms spend half their time in each, lensing of the probe beam is a much smaller issue than with weak nearly resonant probes.  Still, we have measured a density-dependent effect by imaging an atomic cloud loaded into a two-dimensional, blue-detuned optical lattice.  If the cloud is released in a lattice deep enough that tunneling is negligible, it expands along the untrapped direction, while its width remains fixed along the two lattice directions.  The 3D density, and hence the unwanted effect of lensing, drops as a function of evolution time in the 2D lattice, while the true transverse width remains fixed.  The measured transverse width varies by about 10\% (from $5.8~\mu$m to $5.3~\mu$m) when a cloud of initial density $1.5 \times 10^{14}$~cm$^{-3}$ (peak $OT=362$) drops to a final density of $3.7 \times 10^{12}$~cm$^{-3}$ (peak $OT=9$).

In conclusion, we have demonstrated Babinet's principle using a high intensity probe beam passing through a cloud of ultracold atoms.  The technique can be used to image dense atomic clouds.  Like high intensity fluorescent imaging, its Babinet counterpart requires no image processing, but it yields a much higher signal to noise, some 20 times larger than the signal from spontaneously emitted photons collected by the camera. However, it is inconvenient for measuring absolute densities. Also, although density-dependent image distortion is relatively small, it is not wholly avoided in this technique. Finally, unlike low intensity absorption imaging, the technique measures the square of the atomic density integrated along the line of sight and not the line density. This is disadvantageous from the perspective of spatial resolution, but might have a niche application.

{\em Acknowledgements.}  This work was supported by the NSF (PHY 11-02737), the ARO, and DARPA.

\pagebreak


\begin{thebibliography}{99}

\bibitem{Babinet} M. Babinet, C. R. Acad. Sci. \textbf{4}, L638 (1837).

\bibitem{Hecht} E. Hecht and A. Zajac: \emph{Optics} (Addison Wesley; Reading, MA 2001).

\bibitem{DePue} M. T. DePue, S. L. Winoto, D. J. Han, and D. S. Weiss, Opt. Commun. \textbf{180}, 73 (2000).

\bibitem{Nelson} Karl D. Nelson, Xiao Li, and David S. Weiss, Nature Physics \textbf{3}, 556 (2007).

\bibitem{Bakr} W. S. Bakr, A. Peng, M. E. Tai, R. Ma, J. Simon, J. I. Gillen, S. F\"{o}lling, L. Pollet, and M. Greiner, Science \textbf{329}, 547 (2010).

\bibitem{Sherson} Jacob F. Sherson, Christof Weitenberg, Manuel Endres, Marc Cheneau, Immanuel Bloch, and Stefan Kuhr, Nature \textbf{467}, 68 (2010).

\bibitem{Kinoshita} Toshiya Kinoshita and Trevor Wenger and David S. Weiss, Nature \textbf{440}, 900 (2006).

\bibitem{Greiner} Markus Greiner, Olaf Mandel, Tilman Esslinger, Theodor W. H\"{a}nsch, and Immanuel Bloch, Nature \textbf{415}, 39 (2002).

\bibitem{Albiez} Michael Albiez, Rudolf Gati, Jonas F\"{o}lling, Stefan Hunsmann, Matteo Cristiani, and Markus K. Oberthaler, Phys. Rev. Lett \textbf{95}, 010402 (2005).

\bibitem{Greiner2} Markus Greiner, Immanuel Bloch, Olaf Mandel, Theodor W. H\"{a}nsch, and Tilman Esslinger, Phys. Rev. Lett. \textbf{87}, 160405 (2001).

\bibitem{Andrews} M. R. Andrews, M.-O. Mewes, N. J. van Druten, D. S. Durfee, D. M. Kurn, and W. Ketterle, Science \textbf{273}, 84 (1996).

\bibitem{Andrews2} M. R. Andrews, D. M. Kurn, H.-J. Miesner, D. S. Durfee, C. G. Townsend, S. Inouye, and W. Ketterle, Phys. Rev. Lett. \textbf{79}, 553 (19997).

\bibitem{Turner} Lincoln D. Turner, Karl P. Weber, David Paganin, and Robert E. Scholten, Opt. Lett. \textbf{29}, 232 (2004).

\bibitem{Reinhard} Aaron Reinhard, Jean-F\'{e}lix Riou, Laura A. Zundel, David S. Weiss, Shuming Li, Ana Maria Rey, and Rafael Hipolito, Phys. Rev. Lett. \textbf{110}, 033001 (2013).

\bibitem{Labeyrie} G. Labeyrie, G. L. Gattobigio, T. Chaneli\'{e}re, G. L. Lippi, T. Ackemann, R. Kaiser, Eur. Phys. J. D \textbf{41}, 337 (2007).

\bibitem{Reinaudi} T. Reinaudi, T. Lahaye, Z. Wang, and D. Gu\'{e}ry Odelin, Opt. Lett. \textbf{32}, 3143 (2007).

\bibitem{note2}
We have not attempted to image objects with RMS width larger than $180~\mu$m.



\end{thebibliography}
\end{document}